\documentclass[sigconf]{acmart}
\usepackage[utf8]{inputenc}
\usepackage[T1]{fontenc}
\usepackage{textcomp}
\usepackage{xcolor,colortbl}
\usepackage{multirow}
\usepackage{soul}
\usepackage{hyperref} 
\usepackage{subfig}
\usepackage{array}

\AtBeginDocument{%
  \providecommand\BibTeX{{%
    \normalfont B\kern-0.5em{\scshape i\kern-0.25em b}\kern-0.8em\TeX}}}

\newif\ifdraft
\drafttrue

\newcommand{\modi}[1]{\textcolor{black}{#1}}

\copyrightyear{2026}
\acmYear{2026}
\setcopyright{cc}
\setcctype{by-nc-nd}
\acmConference[CHI '26]{Proceedings of the 2026 CHI Conference on Human Factors in Computing Systems}{April 13--17, 2026}{Barcelona, Spain}
\acmBooktitle{Proceedings of the 2026 CHI Conference on Human Factors in Computing Systems (CHI '26), April 13--17, 2026, Barcelona, Spain}
\acmPrice{}
\acmDOI{10.1145/3772318.3791562}
\acmISBN{979-8-4007-2278-3/2026/04}

\begin{document}


\title[Understanding Older Adults' Experiences with Biometric Payment]{``What If My Face Gets Scanned Without Consent'': Older Adults' Experiences with Biometric Payment}

\author{Yue Deng}
\affiliation{%
  \institution{Hong Kong University of Science and Technology}
  \city{Hong Kong}
  \country{China}
}
\affiliation{
  \institution{Max Planck Institute for Security and Privacy}
  \city{Bochum}
  \country{Germany}
}
\email{ydengbi@connect.ust.hk}

\author{Changyang He}
\affiliation{%
  \institution{Max Planck Institute for Security and Privacy}
  \city{Bochum}
  \country{Germany}
}
\email{changyang.he@mpi-sp.org}

\author{Bo Li}
\affiliation{%
 \institution{Hong Kong University of Science and Technology}
  \city{Hong Kong}
  \country{China}
}
\email{bli@cse.ust.hk}

\author{Yixin Zou}
\affiliation{%
  \institution{Max Planck Institute for Security and Privacy}
  \city{Bochum}
  \country{Germany}
}
\email{yixin.zou@mpi-sp.org}

\renewcommand{\shortauthors}{Deng et al.}

\begin{abstract}
Biometric payment, i.e., biometric authentication implemented in digital payment systems, can reduce memory demands and streamline payment for older adults. However, older adults' perceptions and practices regarding biometric payment remain underexplored. We conducted semi-structured interviews with 22 Chinese older adults, including both users and non-users. Participants were motivated to use biometric payment due to convenience and perceived security. However, they also worried about loss of control due to its password-free nature and expressed concerns about biometric data security. Participants also identified desired features for biometric payment, such as lightweight and context-aware cognitive confirmation mechanisms to enhance user control. We outline recommendations for more accessible and informative digital financial services that better support older adults. 
\end{abstract}



\maketitle

\section{Introduction}

Digital payment is now a key part of socioeconomic life, and user interaction in digital payment systems has been an important and well-researched topic in the human-computer interaction (HCI)  literature~\cite{pal2018digital,vines2012cheque,lewis2019follow,li2024beyond,shen2020can,kameswaran2019cash}. Biometric authentication is increasingly implemented in digital payment, commonly referred to as biometric payment \cite{zarco2024comprehensive}. Biometric payment utilizes individuals' unique biometric identification features, such as fingerprints, facial recognition, iris scans, and voice patterns, to allow users to authorize transactions without passwords, PINs or physical cards \cite{
kim2019can}. It enhances convenience by decreasing reliance on passwords and enabling faster transactions~\cite{kumar2009brief}. It also improves security by lowering the risk of card theft and forgery, as biometric data is unique to each individual~\cite{doddipatla2021exploring}. 

\begin{figure*}[htbp]
	\centering
	{\includegraphics[width=1\linewidth]{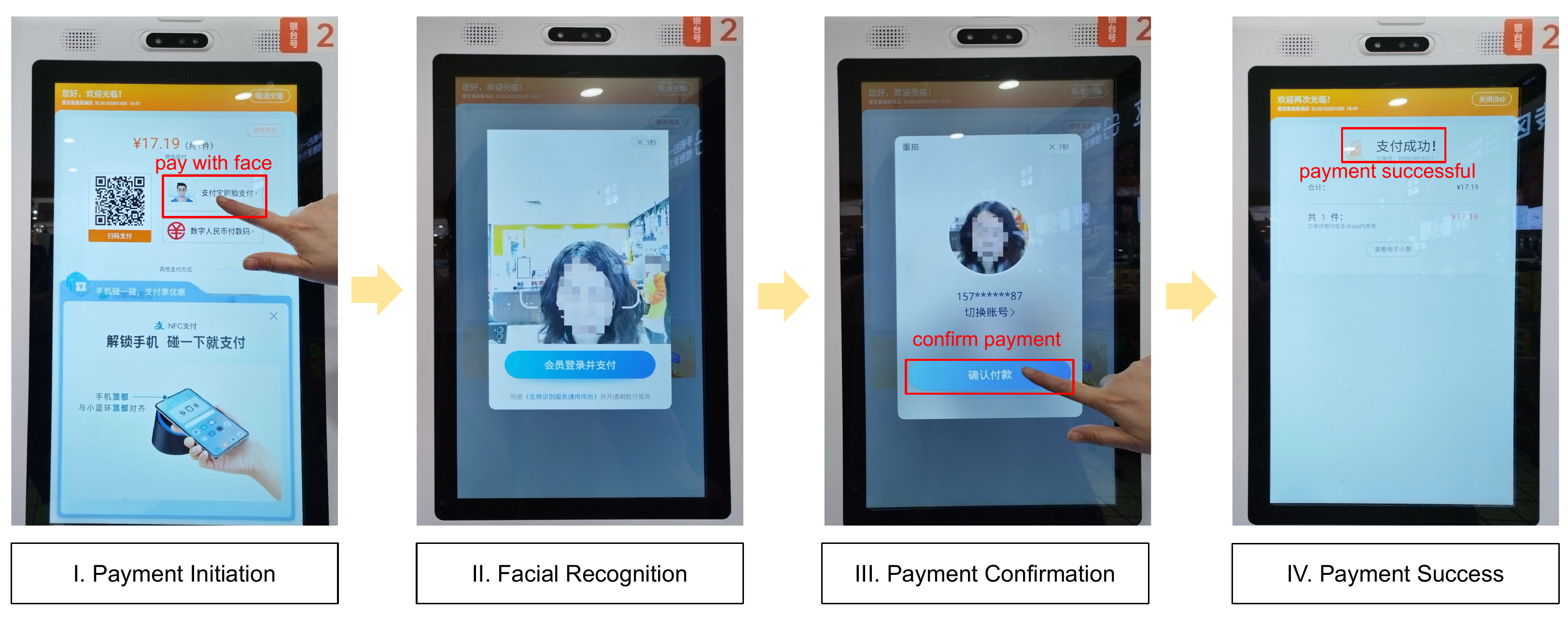}}\\
  \caption{Biometric payment process (an example of an older adult using facial recognition at a supermarket self-checkout kiosk). While interfaces vary across devices and biometric methods, the general process includes: (1) payment initiation (e.g., ``pay with face''), (2) biometric authentication (e.g., face or fingerprint recognition), (3) user confirmation (though some systems proceed directly to payment after recognition without confirmation), and (4) payment result display.}
  \Description{Figure 1: The figure illustrates the biometric payment process at a supermarket self-checkout kiosk, showing four steps: payment initiation, facial recognition, payment confirmation, and payment success.}
        \label{bio-process}

\end{figure*}

Biometric payment has gained global traction. The global facial recognition payment (FRP) market was valued at \$4.62 billion in 2022 \cite{GVR}. Apple Pay employs biometric authentication as its default payment method \cite{url9}. India's BHIM Aadhaar Pay enables transactions via fingerprint verification linked to a national ID \cite{url10}. 
In China, biometric payment is widely adopted in both online transactions through smartphones \cite{yinlian} and in offline environments such as convenience stores, supermarkets, restaurants, and hotels via self-checkout kiosks \cite{zhong2021service,liu2021resistance,gao2025does}. \autoref{bio-process} illustrates the biometric payment process at a supermarket checkout kiosk in China.


As digital payment technologies continue to evolve, it is critical to ensure that older adults are not excluded from digitalization~\cite{omotayo2020digital,das2024design,jin2021too,blanco2015accessible}.
The widespread adoption of digital payment has largely replaced traditional cash-based payment methods, motivating but also forcing older adults to learn unfamiliar payment systems \cite{he2023have,jin2022used}. 
This transition often comes with challenges associated with aging. For example, older adults frequently experience memory challenges, such as forgetting passwords or the steps required to complete digital transactions \cite{abraham2021m,jin2022used}. Additionally, older adults face operational challenges, such as finding it difficult to learn and use digital payment features or apps due to limited experience with digital devices and online interfaces \cite{anwar2024digital}. Biometric payment systems could provide a promising solution for older adults by reducing memory demands and streamlining operational steps. Users do not need to remember or enter passwords, making the payment less reliant on proficiency with smartphone operation, which is senior-friendly \cite{iqbal2020novel}. Furthermore, prior research found that some older adults were able to learn how to use facial recognition payment despite having previously struggled with QR code payment, suggesting biometric payment might be relatively easy to adopt and allow older adults to engage more readily \cite{zhong2022investigating}. 

Previous research on biometric payment usage has primarily focused on the perceptions and practices of the general public (mostly based on younger samples), such as factors that influence customers' intention to use biometric payment \cite{moriuchi2021empirical, hu2023should, zhang2019factors}. Nevertheless, little is known about what motivates older adults to use biometric payment. Therefore, we explore the following research question:

\begin{itemize}
    \item \textbf{RQ1 (motivations)}: What drives older adults to use biometric payment? 
\end{itemize}  

Moreover, biometric technology also carries inherent risks. In India, scammers cloned individuals' Aadhaar-linked biometrics through silicon fingerprints and unauthorized biometric devices, subsequently siphoning money from bank accounts \cite{url14}. In the BioStar 2 incident in Korea, over 27.8 million unencrypted biometric records were exposed due to poor database security \cite{url15}. Existing research also highlights the general public's perceived concerns, such as fears of biometric information leakage \cite{lee2023resistance} and potential financial loss \cite{hu2023should}. However, the perceived concerns of older adults and whether they have adopted protective behaviors to mitigate these risks remain largely underexplored. Understanding what older adults worry about and their current protective measures can help identify their unique needs, reduce their vulnerability to potential risks, and provide valuable insights for better supporting their usage of modern payment services. As such, we also investigate the following research questions:

\begin{itemize}
    \item \textbf{RQ2 (concerns)}: What are older adults’ perceived concerns about using biometric payment?
    \item \textbf{RQ3 (protective behaviors)}: What protective behaviors do older adults adopt when using biometric payment?
\end{itemize}

Additionally, existing research suggests that future studies should go beyond assessing older users' perceptions of biometric technologies and investigate their design expectations, which may help develop systems better suited to their needs \cite{wang2024biometrics}. To this end, we examine the following research question:

\begin{itemize}
    \item \textbf{RQ4 (improvements)}: What improvements do older adults desire in biometric payment?
\end{itemize}




%






To answer these questions, we conducted semi-structured interviews with 22 older adults in China to investigate older adults' perceptions and behaviors around biometric payment. We found that older adults were motivated to use biometric payment due to convenience, security, social influence (e.g., children's recommendations and friends' usage), and prompts from mobile phones and apps.
However, older adults expressed various concerns. For example, they felt a sense of lost control without inputting the password, and they had security and privacy concerns about biometric data leakage and hacking via biometric spoofing. To navigate the trade-off between benefits and risks, older adults adopted a variety of protective behaviors, such as limiting disclosure of biometric data and only using private devices as the biometric payment terminal. Finally, participants suggested several potential improvements, such as integrating lightweight, personalized, or context-aware cognitive confirmation (e.g., contextual voice input) into biometric payment to enhance user control, as well as providing more guidance on using biometric payment. Our work makes the following contributions: 





\begin{itemize}

\item We discover older adults' motivations and concerns regarding biometric payment usage, shedding light on older adults' risk-benefit balance.

\item We unpack older adults' strategic protective behaviors to secure biometric payment use, providing design implications to mediate the interplay between usability and security.


\item We reflect on older adults' desired features in biometric payment, emphasizing the natural alignment between biometric and cognitive authentication processes.


\end{itemize}

\section{Related Work}
\subsection{Biometric Payment Usage}

Biometric payment is defined as ``an authentication system for transactions that relies on the biometric information of each customer''~\cite{zarco2024comprehensive}. Biometric information possesses four key characteristics: universality (everyone has it), distinctiveness (unique to individuals), permanence (stable over time), and collectability (measurable quantitatively) \cite{jain2004introduction}. With identity verification based on unique biological traits, biometric payment presents a promising approach to improving transaction security and convenience~\cite{zarco2024comprehensive}. Advances in biometric technology have enabled authentication using diverse biometric identifiers, such as fingerprints, iris scans, facial recognition, voice, and palm prints, and have facilitated the integration of biometric payment into a broad array of digital finance devices, including personal smartphones and self-service checkout kiosks~\cite{zarco2024comprehensive,kim2019can}. 


A growing body of research has examined users' perceptions of biometric payment. Previous studies have highlighted several perceived benefits of biometric payment. For instance, Hu et al. found that the perceived convenience of biometric payment positively influences perceived value, which in turn builds trust and ultimately drives intention to use \cite{hu2023should}. Similarly, Ogbanufe et al. compared user perceptions of credit cards, credit cards with PIN, and fingerprint biometrics in e-commerce settings, finding that biometric authentication is generally perceived as more secure than the other two methods \cite{ogbanufe2018comparing}. 




However, users also expressed concerns about biometric payment. For example, Mróz-Gorgoń et al. found that fear of losing personal data negatively impacts the perceived safety of biometric payments, with perceived security being a critical factor in shaping attitudes toward biometric systems and predicting behavioral intentions to use them \cite{mroz2022biometrics}. 
Wright emphasized that biometric systems often involve the collection of large amounts of personal data, which heightens concerns about privacy breaches and data leakage \cite{wright2018future}. Beyond security and privacy, factors such as system feature overload, information overload, and technological uncertainty have also been identified as stressors that can lead to user resistance and negative word of mouth \cite{lee2023resistance}. 

\modi{Moreover, several established adoption models have often been used to explain users' attitudes and decisions regarding biometric payment. For example, the Technology Acceptance Model (TAM) highlights perceived usefulness and ease of use as core determinants of technology adoption \cite{davis1989perceived}. Based on it, Zhong et al. showed that perceived enjoyment, coupon availability, and perceived ease of use shape users' biometric payment adoption \cite{zhong2021service}. The Unified Theory of Acceptance and Use of Technology (UTAUT) extends TAM by incorporating factors such as social influence and facilitating conditions \cite{venkatesh2003user}. Zhang et al. applied it and found that social image and environmental factors influence the usage intention of facial recognition payment \cite{zhang2019factors}. The Theory of Planned Behavior (TPB) emphasizes attitudes, subjective norms, and perceived behavioral control in shaping intention \cite{ajzen1991theory}. Hwang et al. found that performance, time, and psychological risks negatively affect attitude toward biometric payment drawn on TPB \cite{hwang2024integrated}.}

The balance between perceived risks and benefits is another defining aspect of biometric payment adoption. For example, Hu et al. explored how Chinese users balance convenience and novelty against potential privacy and financial risks when deciding whether to adopt biometric payment systems \cite{hu2023should}. Similarly, Liu et al. examined privacy-related concerns in facial recognition payment and found that Chinese consumers are often willing to accept privacy risks in exchange for perceived benefits \cite{liu2021resistance}.

Overall, biometric payment offers both benefits, such as enhanced convenience \cite{zhong2022investigating} and security \cite{yu2024acceptance}, but also triggers concerns, such as security and privacy risks \cite{lee2023resistance,zhang2024leaking}. 
However, existing research has predominantly drawn insights from younger generations. Our study contributes an empirical investigation into older adults' perceptions, experiences, and practices regarding biometric payment, providing insights into biometric payment use among this largely understudied population.

\subsection{Older Adults and Digital Payment}





To promote digital inclusion, older adults' adoption of digital payment has been a growing research focus \cite{chawla2018moderating,choudrie2018understanding,zeithaml1987characteristics}. Several studies have examined the factors that motivate older adults to adopt digital payment systems. For instance, Omotayo et al. reported that convenience was the primary reason for older adults using Internet banking \cite{omotayo2020digital}. 
Jin et al. highlighted that some older adults were motivated to adopt new technologies like digital banking as a way to keep their minds sharp, a motivation that is relatively unique to older adults and rarely mentioned by younger users \cite{jin2022used}.

However, older adults also face unique challenges in adopting digital payment systems. They often have lower proficiency in using and learning such technologies. For instance, Chawla et al. indicated that adoption may require spending more time explaining the capabilities of mobile banking or simplifying the interface for older users \cite{chawla2018moderating}. 
Moreover, older adults are often not the primary focus of digital payment systems and have not received sufficient support. For instance, Jin et al. noted that some older adults perceived these systems as being designed primarily for younger users \cite{jin2022used}. Mattila et al. observed that confusing interfaces and complex steps had been identified as significant barriers to Internet banking adoption among older adults, partly due to the lack of comprehensive training on how to use these technologies \cite{mattila2003internet}. 

Older adults expressed significant concerns about digital payment. Trust is one of the most common issues. For example, Azam et al. stated that older adults were skeptical of transactions conducted with no physical interaction, and therefore had little or no trust in them \cite{afshan2013personality}. 
Another concern for older adults is the intangibility of digital banking. Unlike physical banks, which have a physical presence and can provide printed receipts for transactions, digital banks are often perceived as intangible and less concrete \cite{jin2022used}. As suggested by Darch et al., increasing user comfort and enhancing the sense of control should be a priority in influencing older adults' attitudes toward digital payment systems \cite{darch2004investigation}. Third, older adults are concerned about security and privacy risks. Omotayo et al. noted that fear of fraud is a significant reason why many older respondents were reluctant to use Internet banking \cite{omotayo2020digital}. Fourth, older adults may have lower self-efficacy and are less confident in using mobile banking \cite{choudrie2018understanding}. Peral-Peral et al. emphasized that effective training for older adults should not only focus on improving technical skills but also on enhancing their belief in their ability to successfully perform tasks \cite{peral2020self}.
Overall, older adults face unique challenges and hold distinct perceptions regarding digital payments. To what extent they carry over to biometric digital payment remains underexplored. To address this gap, this paper explores older adults' motivations, concerns, protective behaviors, and desired improvements related to biometric payment systems. 

\subsection{Older Adults and Biometric Technology}

A large body of work has examined the performance limitations of biometric technology when used by older adults. 
For example, a fingerprint-based mobile user authentication (MUA) system showed an equal error rate of 30.4–35.8\% for older users \cite{blanco2015accessible}, while achieving an accuracy of 95.7\% for non-elderly users \cite{minaee2019fingernet}. Facial recognition systems achieved a receiver operating characteristic curve of 77.0–83.0\% for older adults \cite{wang2019effect} and an accuracy of 99.2–99.6\% for non-elderly users \cite{yin2018feature}. Wang et al. identified several factors contributing to the lower performance of MUA systems among older adults \cite{wang2024biometrics}. For instance, age-related physiological decline could reduce mobility and dexterity. The issue is exacerbated in older adults with dry or damaged skin \cite{white2011dry}, limited facial expressiveness \cite{corsetti2019face}, or visual impairments \cite{azimi2019effects}. 

Although biometric authentication is often praised for its accessibility and ease of use, these advantages may be compromised by aging and medical complications. \modi{For example, older adults with heart failure may face difficulties when enrolling, authenticating, or verifying through biometric systems, 
because symptoms such as water retention and swelling can cause changes in facial appearance and fingertip characteristics \cite{kowtko2014biometric}.} Galbally et al. found that the quality of fingerprint impressions degrades significantly among adults aged 70 and above, making it the lowest among all age groups considered \cite{galbally2018study}. \modi{A longitudinal study showed that genuine fingerprint match scores tend to decrease as the subject's age increases \cite{yoon2015longitudinal}. Another evidence on fingerprint template aging further demonstrates a statistically significant degradation in recognition performance over time, in datasets spanning up to 21 years \cite{praprotnik2016impact}.} These findings suggest that aging and health-related conditions may present unique recognition challenges for older users, potentially resulting in reduced system performance. When biometric technologies are applied in high-stakes domains such as digital payments, these issues may be further amplified.

Several studies have also explored older adults' perceptions and usage of biometric technologies \cite{kumalasari2024biometric,ahmed2017biometrics,fenuku2024role}. For instance, Kumalasari et al. examined biometric self-authentication for pension fund withdrawals by older adults and found that facilitating conditions and behavioral intention to use were key drivers of continued use, while health-related limitations in vision, hearing, and mobility negatively affected their intention to use \cite{kumalasari2024biometric}. Ahmed et al. investigated biometrics as a substitute for passwords and reported that most older adults perceived biometric options as more secure and providing a better experience than traditional password systems \cite{ahmed2017biometrics}.


In summary, existing work on older adults and biometric technology has primarily focused on performance challenges or perceptions of biometric use in non-payment contexts. However, little is known about how older adults perceive and engage with biometric technologies when used specifically for digital payments. Therefore, our research extends this line of work by examining older adults’ perceptions and practices in the context of biometric payment systems.

\section{Methods}
We opted for semi-structured interviews due to the exploratory nature of our study. The interview approach allowed us to gather in-depth data on participants' perceptions and experiences related to biometric payment, while enabling follow-up questions and the flexibility for participants to share additional insights. Specifically, we interviewed 22 older adults in July and August 2024 to explore older adults' motivations, concerns, protective behaviors, and expected improvements in biometric payment. 
\subsection{Participant Recruitment}

\textbf{Eligibility.} 
The eligibility criteria are twofold. First, participants should be above 50 years old -- a criterion that mirrors the legal earliest retirement age in China \cite{retiree} and is consistent with prior research involving older adults \cite{he2023have,wan2019appmod,herbert2022fast,mcdonald2023don}. Second, the participant should have experience using mobile payment, which ensures they are relatively familiar with mobile phones and payment services and can provide reasonable responses about biometric payment that are not fraught with misconceptions.
We included both participants with and without experience using biometric payments, as this approach provides a more comprehensive and nuanced understanding of user perspectives. For instance, non-users can offer insights into the concerns that prevent adoption, while users can share their concerns after adoption.



\textbf{Recruitment channels.} Initially, we intended to hand out flyers, but we discovered that many older adults were hesitant to discuss sensitive topics, particularly those related to payment. Therefore, instead of directly distributing flyers, we visited places where older adults frequently gather such as senior activity centers and senior residential communities, a commonly used approach for engaging seniors \cite{hornung2017navigating}. We started casual conversations with them to introduce ourselves, explain the purpose and value of the study, and outline participation expectations, thereby inviting individuals to take part in the research and establishing rapport. Additionally, to broaden our recruitment scope, we advertised our study through posters on social media platforms widely used by the older adult population in China, such as Douyin and WeChat groups. Prior to advertising, we obtained permission from group moderators. In our advertisement poster, we included the contact information of the first author. Since filling out a screening survey might have been time-consuming and burdensome for older adults, we assessed their eligibility during casual conversations, whether conducted in person or online. \modi{During the casual conversations, we primarily focused on confirming the participant's age and whether they had used mobile payment, along with a few background questions such as which payment methods they use. This screening took only a few minutes. The online and offline procedures were the same, with online screening conducted via calls or text and offline screening conducted in person.}


Moreover, we adopted a snowball sampling method \modi{\cite{biernacki1981snowball}}, which potentially made it easier to gain the trust of older participants. To diversify participants' demographic backgrounds, we initially reached out to older adults in three cities with different tiers in China\footnote{\modi{Following widely used market-based classifications of Chinese cities, cities across different tiers often reflect variations in the concentration of financial resources, urban connectivity, population activity, lifestyle diversity, and future adaptability \cite{tier}. Tier 1 cities are generally regarded as the country’s leading metropolitan centers, while New Tier 1, Tier 2, Tier 3, and Tier 4 cities are often described as occupying progressively lower positions in China’s urban hierarchy, typically functioning as emerging national hubs, strong regional centers, mid-level local hubs, and small local cities, respectively.}} (i.e., Guangzhou, Chongqing, and Beihai) and encouraged them to refer others. \modi{Our recruitment was conducted on a rolling basis. As participants joined, we interviewed them and analyzed the emerging data. When we reached 17 participants, the responses from older adults were approaching saturation. To confirm that saturation had truly been achieved, we recruited an additional five participants, bringing the final sample to 22.}

\textbf{Demographics.} Table \ref{basic information} describes the background of the interviewed participants. Their ages ranged from 50 to 73. Seventeen participants are women and five are men. 
They came from seven cities spanning four tiers (i.e., Tier 1: Guangzhou and Shenzhen, New Tier 1: Chongqing and Chengdu, Tier 3: Handan, Tier 4: Jilin and Beihai). Among all 22 participants, 13 had used biometric payment, and the remaining had not. The main biometric payment methods utilized by older adults were face recognition ($n$$=$$10$) and fingerprint recognition ($n$$=$$9$). The majority of biometric payment terminals used were smartphones ($n$$=$$13$). Among them, many mentioned employing facial or fingerprint recognition to complete online shopping purchases and bank transfers. Three participants also recounted using facial recognition at checkout terminals in physical stores. One older adult mentioned using facial recognition for automatic fare deduction when entering the subway.

\begin{table}[h]
\caption{Basic Information of Interviewed Seniors. (CQ-Chongqing, BH-Beihai, GZ-Guangzhou, HD-Handan, SZ-Shenzhen, JL-Jilin, CD-Chengdu)}
\Description{Table 1: The table presents demographic information of interviewed seniors, including ID, age, sex, city, bio-payment methods (face or fingerprint), and devices used (such as smartphone, checkout terminal, or fare gate).}
\label{basic information}
\rowcolors{2}{gray!20}{white}
\begin{tabular}{lrllll}
\hline
\textbf{ID} & \multicolumn{1}{l}{\textbf{Age}} & \textbf{Sex} & \textbf{City} & \textbf{Bio-payment}                                                    & \textbf{Devices}                                    \\ \hline
P01         & 62                               & M               & CQ     & face                                                                      & smartphone                                                                 \\
P02         & 61                               & F               & CQ     & /                                                                                    & /                                                                     \\
P03         & 55                               & F               & BH        & /                                                                                    & /                                                                     \\
P04         & 52                               & F               & CQ     & face                                                                      & smartphone                                                                 \\
P05         & 60                               & F               & BH        & /                                                                                    & /                                                                     \\
P06         & 61                               & F               & GZ     & \begin{tabular}[c]{@{}l@{}}face, fingerprint\end{tabular} & \begin{tabular}[c]{@{}l@{}}fare gate, \\ smartphone\end{tabular}
\\
P07         & 52                               & F               & CQ     & face                                                                      & smartphone                                                                 \\
P08         & 62                               & F               & CQ     & /                                                                      & /                                                                     \\
P09         & 70                               & M               & CQ     & \begin{tabular}[c]{@{}l@{}}face, fingerprint \end{tabular} & smartphone                                                                 \\
P10         & 50                               & M               & HD         & \begin{tabular}[c]{@{}l@{}}face, fingerprint \end{tabular} & smartphone                                                                 \\
P11         & 67                               & F               & GZ     & /                                                                                    & /                                                                     \\
P12         & 60                               & M               & SZ      & fingerprint                                                               & smartphone                                                                 \\
P13         & 67                               & F               & CQ     & /                                                                                    & /                                                                     \\
P14         & 73                               & F               & JL         & /                                                                                    & /                                                                     \\
P15         & 65                               & F               & BH        & \begin{tabular}[c]{@{}l@{}}face, fingerprint \end{tabular}  & \begin{tabular}[c]{@{}l@{}}checkout terminal, \\ smartphone\end{tabular} \\
P16         & 52                               & F               & CQ     & /                                                                                    & /                                                                     \\
P17         & 71                               & F               & CQ     & /                                                               & /                                                                     \\
P18         & 50                               & F               & CD       & face, fingerprint                                                                                      & smartphone                                                                     \\
P19         & 60                               & F               & SZ      & face                                                                                     & \begin{tabular}[c]{@{}l@{}}checkout terminal, \\ smartphone\end{tabular}                                                                     \\
P20         & 51                               & M               & CQ     & fingerprint                                                                                     & smartphone                                                                      \\
P21         & 50                               & F               & CQ     & fingerprint                                                                                     & smartphone                                                                      \\
P22         & 55                               & F               & BH        & face, fingerprint                                                                                     & \begin{tabular}[c]{@{}l@{}}checkout terminal, \\ smartphone\end{tabular}                           \\ \hline                           
\end{tabular}
\end{table}

\subsection{Semi-structured Interview}
\textbf{Interview guide.} To develop our interview guide, we personally experienced a variety of biometric payment methods and devices in China to gain a deeper understanding of their usage. This included using facial recognition and palm recognition payments at public terminals such as checkout kiosks, as well as fingerprint and facial recognition payments on private devices such as smartphones.
Based on these investigations and our research questions, we designed an interview guide, as shown in Figure \ref{FIG: interview guide}, with detailed questions provided in Appendix \ref{interview guide}. Our research team iteratively reviewed the interview questions to ensure appropriate coverage of relevant aspects of biometric payment.

Specifically, to establish a basic understanding, we first asked older adults about their familiarity with biometric payment, such as whether they had heard of it and whether they had used it before. For those who had used biometric payment, we further inquired about their specific usage, such as the biometric payment methods they adopted and the devices they used. Next, we explored general perceptions of biometric payment. For users, we focused on their usage experiences, feelings, and considerations. For non-users, we delved into their reasons for not using biometric payment, their willingness to try it, and other related considerations. After that, we probed the users' motivations for using biometric payment, such as their reasons for using them and comparisons with other payment methods.
In the next section, we asked all participants targeted questions about their concerns related to biometric payment, including the difficulties they faced, the sources of their concerns, their trust in various stakeholders, and so on. Then, we also inquired whether they had adopted specific measures to protect their information or money. Finally, we asked participants about their desired improvements for biometric payment, focusing on the features and designs they believed could alleviate their concerns.

\begin{figure}[htbp]
	\centering
	{\includegraphics[width=1\columnwidth]{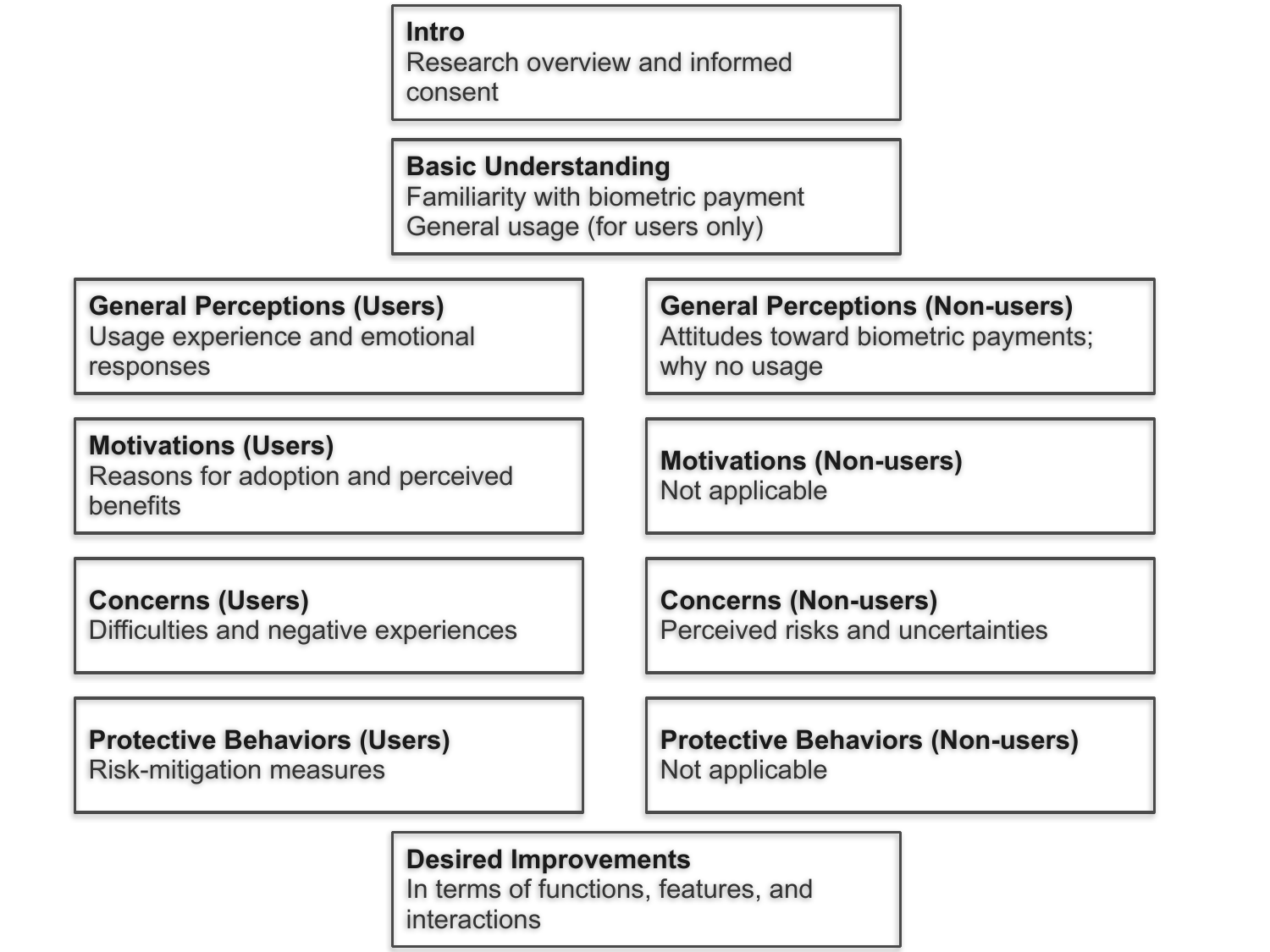}}\\
  \caption{Overview of the interview process.}
  \Description{Figure 2: The figure shows a flowchart of the interview process, starting with introduction and consent, followed by sections on basic understanding, perceptions, motivations, concerns, protective behaviors, desired improvements, and ending with feedback. Paths are shown for both users and non-users of biometric payment.}
  \label{FIG: interview guide}
\end{figure}

\textbf{Pilot interview.}
Prior to conducting the formal interviews, we conducted five pilot interviews to evaluate the clarity of the interview questions and the interview duration. The recruitment criteria for the pilot interviews were identical to those for the formal interviews. Based on feedback from the pilot participants, we made minor adjustments to the order of questions to improve the interview flow and refined some phrasing to enhance comprehension for older adults. For example, we opted to use ``facial recognition payment'' or ``fingerprint recognition payment'' in the interview questions, as some older adults were less familiar with the term ``biometric payment.'' By the fifth pilot interview, no participants expressed any confusion, and the interview guide was finalized. As the changes were minor to improve clarity and comprehension, the data from the pilot interviews were included in the final coding and analysis, which is a common practice when pilot modifications are minimal \cite{haney2021s}.

\textbf{Interview procedure.}
\modi{At the beginning of the interview, we showed the consent form with participants and further explained it such as the research goal, the interview process and duration, potential benefits and risks, and how the interview data would be used. Participants could ask any questions about the study, and audio-recorded oral consent was then obtained.\footnote{According to guidance from our university’s Institutional Review Board (IRB), when conducting research where seeking written consent is not practical or too sensitive, audio-recorded oral consent might be less of a privacy risk than written consent and can be considered as an alternative. In our study, older adult participants are particularly sensitive to signing formal documents, especially in discussions related to payment.}} Although we had an interview guide, the interviews were conducted in a semi-structured format, allowing flexibility to follow up on specific questions or topics in depth to elicit unexpected but relevant responses. Additionally, we skipped some questions if the participant had already addressed the information in their earlier answers. At the end of the interview, we encouraged the participant to provide any additional insights we might have overlooked. 

Based on participants' preferences, we conducted the interviews either in person or via voice calls. Among the 22 interviews, three were conducted in person, 17 via voice calls, and two started in person and were later continued via voice calls. Some participants recruited offline opted to conduct the interview via voice calls. \modi{The average duration of interviews is 28.43 minutes.}

With the participant's consent, all interviews were audio-recorded and transcribed using Feishu Minutes, a tool that converts audio into searchable and interactive text \cite{feishu}. After completing the interview, we thanked older adults for their efforts and provided them with a 50-yuan (US\$6.9) compensation.\footnote {The compensation was determined with the aim of exceeding the hourly minimum wage of the province where each participant's city is located. (Chongqing-21.0 yuan, Beihai-20.1 yuan, Guangzhou-22.2 yuan, Handan-22.0 yuan, Shenzhen-22.2 yuan, Jilin-21.0 yuan, Chengdu-22.0 yuan) \cite{url12}).} 


\subsection{Coding and Analysis}
We conducted an inductive thematic analysis \cite{braun2006using} of our interview transcripts. The first author manually reviewed transcripts to correct potential transcription errors. Then, two authors individually went through these transcripts using an open coding method \cite{corbin2014basics}, allowing codes to naturally emerge to capture older adults’ biometric payment usage. Examples of initial codes included \textit{biometric payment lacks a confirmation process}. Next, the researchers iteratively compared and discussed their codes, resolving disagreements through multiple rounds of meetings. We further used affinity diagramming \cite{muller2014curiosity} to group codes with similar meanings into higher-level themes and excluded codes less related to older adults' biometric payment usage (e.g., \textit{young people could easily accept new things compared with old people}). Ultimately, our codebook comprised 21 themes (e.g., \textit{security and privacy concerns}), 60 sub-themes (e.g., \textit{fear of biometric data leakage}), and 244 codes.

Since our coding process involved multiple iterative rounds of coding, discussion, and refinement to generate themes, codes here served as a process rather than a final product, which is one of the reasons for not seeking inter-rater reliability (IRR) according to McDonald et al.~\cite{mcdonald2019reliability}. 
\modi{All quotes were initially translated from Chinese to English using Google Translate. The first author then refined the translations to ensure accuracy.}

\subsection{Ethical Considerations}

\modi{We followed our university’s Policy on the Ethical Conduct of Research Involving Human Participants, which is grounded in three overarching ethical principles: beneficence, justice, and respect for persons.\footnote{\url{https://vprd.hkust.edu.hk/policies-compliance/policies-guidelines/human-participants}}} Our research received approval from our institution's Institutional Review Board (IRB). All audio recordings were transcribed, and the resulting transcripts were securely stored without any personal identifiers. Each participant's transcript was labeled with generic identifiers such as ``P01'' and access to the data was strictly limited to the research team. Participation was entirely voluntary. \modi{Prior to obtaining consent, participants were thoroughly informed about the study's purpose, data use, potential benefits and risks, and could ask any questions they were concerned about. Participants were informed that they could decline to participate without any repercussions.} They were further assured of their right to skip any question for any reason or withdraw from the interview at any point without impacting the overall process. To help alleviate stress or concerns about difficulty in answering, we reassured participants that the research questions were focused on their personal thoughts and everyday experiences and that there were no right or wrong answers. 

\subsection{Limitations}

Our study has several limitations common to interview-based research, including the potential for over-reporting, under-reporting, recall bias, social desirability bias, and sampling bias \cite{kostan2024exploring,haney2021s,ray2021older}. Moreover, we did not include older adults who have never used mobile payments, so their unique perceptions and behaviors were not captured in this study. Additionally, our sample consists predominantly of female participants, which may introduce a gender bias in the findings. Furthermore, our study focuses on biometric payments among older adults in China, which limits the generalizability of the findings to other cultural and social contexts. Future research could explore and compare the perspectives of older adults in different regions to identify the impact of cultural factors on their perceptions and behaviors. \modi{In addition, though our analysis does not deliberately control for specific biometric payment methods, it largely reflects older adults' perceptions and practices of fingerprint and facial recognition payments as the dominant biometric payment forms in the Chinese market. Newer biometric options, such as iris or palm recognition payments, are less familiar to older adults and may introduce additional cognitive burden if specially introduced. As these emerging technologies evolve and begin to appear in older users' daily lives, future research could explore them to assess whether they reveal new insights.}


\section{Findings}
In this section, we present seniors' motivations for using biometric payment in Sec. \ref{motivations}, including security, convenience, social influence, and facilitating conditions. Following that, Sec. \ref{concerns} captures four types of concerns, including lack of control, security and privacy concerns, lack of knowledge about how to use biometric payment, and difficulties adapting to different payment scenarios. We also examine older adults' protective measures when using biometric payment in Sec. \ref{protective behaviors}. Lastly, we present findings regarding their desired improvements for biometric payment in Sec. \ref{improvements}.

\subsection{Motivation for Using Biometric Payment} \label{motivations}

\subsubsection{Convenience.}
Convenience is a major driver for using biometric payment. Older adults appreciate that using biometric payments reduces operational burdens and saves time during transactions. For instance, P09 (M, 70, user) mentioned that biometric payments were faster and eliminated the worry of entering the wrong password, \textit{``Fingerprints are more convenient. With a fingerprint, you just press it and you're done.''} 
Another advantage highlighted by participants is that biometric payments enable shopping without the need to carry a mobile phone. 
As P19 (F, 60, user) stated,
\begin{quote}
    \textit{``It's especially convenient. I really like biometric payments...wherever you go, even if your phone dies, you can still pay with facial recognition.''}
\end{quote}

Furthermore, even when participants are aware of the potential risks of biometric payment, they are still willing to accept them in exchange for convenience. P15 (F, 65, user) reflected on this trade-off,
\begin{quote}
    \textit{``Financial loss is one aspect, and on the other, someone might take your biometric data and use it, resulting in irreparable damage \dots I was aware of these risks from the very beginning. However, when you choose convenience, you simply can't have it both ways.''}
\end{quote} 

\subsubsection{Perception of higher security.} One key motivation is the perception of enhanced security. They believe that biometric methods effectively prevent password exposure during transactions---a significant advantage over traditional password-based systems. P20 (M, 51, user) mentioned, 
\begin{quote}
    \textit{``I don't think the password is necessarily secure. If you're under surveillance, your password might be exposed. But with fingerprint-based payments, there's no risk of such surveillance.''} 
\end{quote}



\subsubsection{Social influence.} Recommendations from seniors' children can motivate the elderly to use biometric payments. For instance, P09 (M, 70, user) explained, \textit{``It was my son and daughter-in-law who taught me how to use it and helped me set it up... They said it was more convenient, so I used it.''} 

Beyond recommendations from children, seeing people nearby use biometric payments also influences older adults to adopt the technology. As P18 (F, 50, user) pointed out, \textit{``I didn't really overthink it. Many people were using it, so it just felt natural to set up. I saw how convenient it was for others, figured it was safe, and decided to go for it.''} 


\subsubsection{Reminders and incentives from mobile apps.} 
Adoption of biometric payment can also originate from supportive infrastructure. Many older adults have found that their mobile apps automatically prompt them to activate biometric payment, thereby reducing the burden of having to find the activation interface themselves. 
For example, P20 (M, 51, user) said, \textit{``My mobile banking app popped up with the fingerprint option, so I just pressed it.''} However, this ease of activation can sometimes result in unintentional sign-ups. As P06 (F, 61, user) mentioned, \textit{``I activated it following the instructions from [a bank name]. I just did it without really understanding which function I was pressing.''} 



Some apps also offer discounts for using biometric payments, 
which further encourages older users to adopt the technology. P21 (F, 50, user) recalled, \textit{``A cashier told me, `If you use Alipay's facial recognition payment, you'll get a discount.' That got me interested, so I decided to try it out.''}

Additionally, biometric payment systems backed by reputable state-owned enterprises can help alleviate seniors' concerns about security. P06 (F, 61, user) remarked, \textit{``I have concerns about biometric payment in other organizations, but I'm not worried about the subway's biometric payment because it's backed by a state-owned enterprise.''} 

\subsection{Concerns Regarding Biometric Payment} \label{concerns}

\subsubsection{Lack of Knowledge}
\label{Lack of Usage Knowledge}
A concern among older adults is their lack of knowledge about using biometric payment. They worry that this lack of understanding might lead to operational errors and potential losses. For example, P06 (F, 61, user) expressed a fear of exploring unfamiliar systems, \textit{``Trying new things feels like exploring the unknown. If no one explains it to us, we're afraid of making mistakes or causing problems.''} Similarly, P17 (F, 71, non-user) mentioned a fear of monetary loss, \textit{``The main worry is that if you make a small mistake on biometric payment, the money might go somewhere else.''} 

Furthermore, older adults' limited knowledge of how to use biometric payment can restrict their autonomy. P18 (F, 50, user) explained that she was willing to use it but had not been shown how, \textit{``My colleague uses face scanning to enter the subway, but he hasn’t taught me yet. I still use QR code payment.''} In addition to learning how to pay through biometric payment, knowing how to cancel it is also important. P21 (F, 50, user) said, \textit{``I signed up for it, but I can’t figure out how to cancel.''} 
Therefore, older adults who have already activated biometric payment but do not know how to cancel it are forced to continue using it. 


Additionally, when older adults are unfamiliar with biometric payment and are unaware of the actual scope of biometric payment usage, they may hold misconceptions that do not reflect reality. For example, P12 (M, 60, user) commented, \textit{``It seems there's a restriction, right? You can only use it at that one store, rather than any other stores?''} In reality, biometric payments can be used at any location equipped with the necessary biometric payment devices, without being limited to a specific store. 

\subsubsection{Lack of Control}
\label{lack of control}
Since biometric payment systems can verify and process transactions using biometric data without password input, some older adults felt that there was a lack of cognitive payment confirmation and worried that payments might be made without their genuine intent. P03 (F, 55, non-user) expressed, 
\begin{quote}
    \textit{``It just feels unsafe. What if my face gets scanned when I don't need it to? For example, I'm not trying to make a payment, but it still scans my face and processes a payment without my consent.''}
\end{quote} 

For some, using passwords is a deeply ingrained habit, a step that serves as a cognitive confirmation of their consent for a payment. When this step is removed and replaced by biometric methods like facial recognition or fingerprint scanning, they feel that the transaction bypasses their conscious approval. As P14 (F, 73, non-user) explained, \begin{quote}
\textit{``Using a password feels like putting a lock on something. With fingerprints or facial recognition, it just feels less secure. Logically, it could be the same, but I can't seem to change my perspective.''}
\end{quote} 


Moreover, the declining abilities of older adults further exacerbate their sense of lack of control. For instance, P06 (F, 61, user) mentioned their deteriorating vision, which might lead to unintended actions,\textit{``Our eyesight deteriorates with age. For example, while scrolling through TikTok or other apps, I might accidentally swipe to open my banking app, and it scans my face when I don't want it to. If someone is nearby, they could see it.''} 

The fear of losing control can partly explain why some older adults find fingerprint payment more proactive and reassuring compared to facial recognition payment. Fingerprint payment requires a deliberate action, such as raising a hand, which provides a sense of cognitive confirmation. In contrast, facial recognition feels more passive and automatic. As P12 (M, 60, user) explained, \textit{``With fingerprints, you see the payment prompt, and you actively decide to press your finger. It's not as prone to misunderstandings. You won't extend your hand unless you've agreed to the payment first.''} 

\subsubsection{Security and Privacy Concerns} \label{sp concerns}
\textbf{Unauthorized biometric data collection.}
Older adults are concerned about the possibility of their biometric data being collected illegally. 
For example, P08 (F, 62, non-user) expressed worry about her fingerprints being extracted without authorization, \textit{``I've seen short videos saying that home door locks now use fingerprint recognition. With just a spray of powder, your fingerprint becomes crystal clear, and someone could even create a mold of it.''}

\textbf{Biometric data leakage.}
Elderly users are also very concerned that their biometric data might be leaked after collection. Many recall how their phone numbers have been exposed and worry that their biometric information could be treated in the same way. For example, P18 (F, 50, user) remarked, \textit{``Once biometric data is collected, there's a risk it might be leaked just like our phone numbers. I often get all kinds of telemarketing calls. Essentially, it’s because the data protection isn’t done properly. I'm afraid that once others get hold of our biometric information, they'll sell it everywhere.''} 


Additionally, some older adults perceived their personal information as interconnected, leading to concerns that if their biometric data were compromised, other sensitive information might also be exposed. 
P14 (F, 73, non-user) elaborated, \textit{``It's all linked together. For example, your bank cards, credit cards, health insurance, and social security cards, all of these could be exposed. That's why I don't want to enable biometric payments.''}

\textbf{Hacking.}
Older adults are fearful of hacking. P18 (F, 50, user) explained, \textit{``I worry that some criminals might use viruses or other means to exploit my biometric information and then hack into my account.''} 
Furthermore, given that some participants believe that their biometric data is interconnected with other personal information shown (Sec. \ref{sp concerns}), they worry that hackers could sell or trade all of this data on the black market. P15 (F, 65, user) explained, \textit{``I'm worried that my identity, phone number, and image data could be bought and sold by hackers.''}

In addition, hacking through biometric spoofing is one of the main concerns among older adults. 
P04 (F, 52, user) shared, \textit{``One of my friends had his overdraft card stolen. He fell victim to facial recognition fraud when a scammer video-called him, without much thought, he answered, and the scammer used his video to spoof his face. He ended up losing over 90,000 yuan.''} Furthermore, older adults' concerns about hacking via biometric spoofing in payment systems are strongly influenced by their experiences with biometric authentication in non-payment settings. They worry that if biometric authentication can be easily hacked in other scenarios, the same vulnerabilities may be exploited to bypass payment security. P06 (F, 61, user) explained,
\begin{quote}
\textit{``At my workplace, we also use facial recognition. When someone is supposed to be at a project site, they might use an A4-sized photo of their face to scan instead of showing up in person. The machine sometimes accepts it. If such flaws exist, how can it be trusted for payments?''}
\end{quote} She also believed that the reliability of facial recognition technology in non-payment and payment contexts is inherently the same, \textit{``When it recognizes someone for attendance, it works the same way as it does for payments.''} Although this scenario may not be accurate -- different biometric authentication applications typically operate independently -- it still fuels concerns among participants.

\textbf{Identity theft.}
Older adults are also deeply worried about identity theft. They fear that their biometric data could be illegally used to exploit their credit.
For example, P13 (F, 67, non-user) expressed that she's concerned that her biometric data might be used to apply for illegal loans,
\begin{quote}
    \textit{``My concern is that someone might take my biometric information and use it to secure illegal loans. That would be a big problem because I have no idea how much they could loan, maybe millions.''}
\end{quote} 

\subsubsection{Low Adaptability}
\label{adaptability}
Older adults worry that biometric recognition systems lack adaptability when used for payment. First, they are concerned that \textbf{aging-related changes} in their biometric features will lead to frequent identification failures, making the supposed convenience of biometric payment inconvenient. As P06 (F, 61, user) put it, 
\begin{quote}
    \textit{``Now I'm older, my eyelids droop, and my face puffs up...I signed up for facial recognition before 55, and now five years have passed. It used to recognize me instantly, but these days, any slight change sets off extra verification steps.''}
\end{quote}

Second, they are concerned that \textbf{temporary changes} in their biometric features caused by daily activities, such as swelling, moisture, or injury, may also result in failed recognition. P06 (F, 61, user) also shared, \textit{``I registered my fingerprint when my hands were dry, but if I wash dishes and my fingers are damp, the system won’t recognize me. If I carry luggage, my fingers get swollen from the pressure and it fails again. Or if I cut my finger while slicing fruit, it won’t recognize me either.''}

Lastly, another concern for older adults is related to \textbf{environmental conditions}, such as poor lighting or having multiple people in the camera’s view can cause recognition to fail. As P15 (F, 65, user) noted, \textit{``At malls or supermarkets, sometimes the lighting or screen is a bit dim, or there are people standing next to or behind you, and it won’t recognize you.''} 



\subsection{Protective Behaviors} \label{protective behaviors}


\textbf{Limiting disclosure of biometric data.}
When distrusting certain institutions or technologies, some older adults choose to protect their privacy by refusing or reducing the disclosure of personal biometric data in certain contexts. They do not want their biometric information widely collected or stored due to fears of potential leaks or misuse. For instance, P18 (F, 50, user) provides her biometric information in a payment system, yet she deliberately avoids disclosing it in other contexts to prevent further unauthorized use or storage. She said, \textit{``For residential gate access control, the property management company will collect our facial data (so that residents can enter without access cards, passwords, etc.) I don’t trust them. I’m worried my information will be leaked, so I never use their face-recognition system.''} Another example is P04 (F, 52, user), who avoids video calls to prevent her facial data or other sensitive information from being captured and exploited. 
She explained, \textit{``I don’t do video calls with people I’m not familiar with.''} 

\textbf{Only using private devices as the biometric payment terminal.}
Older adults mentioned that there are various types of biometric payment devices, including both private devices, such as smartphones, and public devices like checkout kiosks. Avoidance of public devices 
can happen as a result of perceived lack of control over public devices and uncertainty about potential security or privacy risks associated with these devices. Specifically, P18 (F, 50, user) stated, \textit{``I haven’t used the merchant’s checkout kiosk yet because I can control my phone to avoid downloading random apps and prevent malware. However, I can’t be sure about the security of the merchant’s devices, such as whether they might have viruses or if they could collect my information.''}

\textbf{Only enabling specific biometric payment applications.}
Many applications nowadays offer biometric payment functions, but some elderly individuals selectively activate biometric payment on specific and trusted apps. They intentionally avoid enabling biometric payment across multiple platforms, believing that restricting biometric payment to one carefully chosen app could reduce the likelihood of accidental exposure to threats such as phishing and data leakage. P06 (F, 61, user) activated biometric payment only on her mobile banking app based on her trust in banks, \textit{``Alipay and WeChat also have facial recognition, but I didn't activate it. I only activated the one with mobile banking. At our age, it's easy to fall for phishing or accidentally make mistakes.''} 
P01 (M, 62, user) perceived Alipay as more secure due to its relatively singular focus on payment functionality,  \textit{``I'm afraid there are too many channels information can leak, so I only activate one. Alipay feels trustworthy. It's dedicated mainly to payments, so the function is simpler and feels safer. Unlike WeChat, which includes chatting and transfers altogether.''}

\textbf{Selective use of biometric payment based on payment amounts.} Flexibly adjusting payment methods according to the transaction amount is also one of the protective behaviors. For smaller payments, biometric payment is often favored due to its convenience. In contrast, password entry is typically used for larger transactions, which provides greater psychological security and reduces potentially unexpected risks associated with biometric payments. P04 (F, 52, user) stated, \textit{``For large amounts, I use passwords, and for small amounts, I use facial payment. I set it up so that once it exceeds a certain amount, I must enter the password.''} The possible reason for this behavior is that some older adults perceive passwords as inherently safer than biometric methods (Sec. \ref{lack of control}).

\textbf{Using a dedicated low-balance card.}
As many platforms nowadays require users to link bank cards online for spending, one protective behavior adopted by older adults is to intentionally link only a dedicated card, usually with a low balance. This approach allows them to concentrate potential risk on a single, non-primary account, thereby avoiding exposure of their main financial accounts. For example, 
P15 (F, 65, user) described her approach of maintaining a consistent balance on a designated card, \textit{``I always use the same card online, and I keep roughly the same amount of money on it. After I spend it, I top it up again. That card is just for spending.''}

\textbf{Enabling instant bank payment notifications.}
A protective behavior to enhance a sense of financial control is enabling real-time notifications for all bank transactions, 
so that one can monitor activity and detect suspicious payments promptly. P09 (M, 70, user) shared, \textit{``I went to the bank and specifically requested that they send me a notification every time a payment is made...I’ve set it so that even if I pay just 5 or 3 yuan, I get a message right away.''} 

\subsection{Desired Improvements for Biometric Payment} \label{improvements}

\subsubsection{Enhancing Cognitive Confirmation for Biometric Payment}
\label{cognitive confirmation}
Older adults expressed concerns about the lack of control in biometric payment, which allows transactions without password input, leading to a perceived absence of cognitive confirmation and fears of unintended payments (Section \ref{lack of control}). To address this, older adults expressed a desire for improved recognition mechanisms that combine cognitive input with biometric recognition to strengthen the payment confirmation process.

\textbf{Lightweight behavioral verification plus biometric recognition.} Some participants suggested introducing simple behavioral actions into the biometric payment process to reinforce a sense of intentionality and user involvement. One suggestion is system-prompted facial actions. A participant proposed that the system could randomly prompt users to perform simple facial movements, such as blinking or turning their heads, to verify that the user is consciously participating in the transaction. As P18 (F, 50, user) described, 
\begin{quote}
    \textit{``In the future, biometric payment shouldn't be completed just by scanning a face. There should be an extra step, like blinking or turning your head, to verify that it’s really me, and that I’m responding voluntarily. That would make it feel more secure. These simple actions could be different each time, adding some randomness. They're just easy actions and wouldn’t create extra burden for the user.''}
\end{quote}This approach ensures that the system is interacting with a real rather than a spoofed image or an unintended action, which may enhance the user’s perceived control.

Another idea is user-predefined gestures. One older adult envisioned a more personalized method, where each user can create and update their own hand gesture, similar to a visual password. As P21 (F, 50, user) shared, \textit{``I can set the `Yeah' gesture, someone else might choose `OK'. Everyone has their own gestures, like a password that can be changed.''} 

\textbf{Contextual voice input plus biometric recognition.} Another older adult suggested enhancing biometric payment by combining it with voice input. The idea is 
to incorporate context-relevant semantic content, such as stating the purchase amount, as a situational form of identity verification. Specifically, P11 (F, 67, non-user) explained,
\begin{quote}
    \textit{``I think adding my voice could help. Using both facial and voice recognition might feel more secure. For example, if I'm buying something that costs 105 yuan, I could say, `I'm [nickname], I want to pay 105.' I don't need to say my full name. And even if someone steals my voice, they wouldn't be able to replicate it exactly.''}
\end{quote} By incorporating contextual cues that match the specific transaction scenario, the authentication becomes more dynamic and makes it more difficult to spoof through voice cloning alone.

\textbf{Custom visual patterns plus biometric recognition.} In addition to biometric recognition, drawing a user-defined ``visual code'' on the screen of biometric payment terminals during the payment process, such as using simple shapes or characters, was suggested by an older adult as another form of cognitive input. As P11 (F, 67, non-user) explained, \textit{``I could draw my own code. For example, simple shapes like triangles, circles, or even letters would work. After a while, I could switch it to a different one, not changing it every day, but maybe from time to time.''}
\subsubsection{Providing Biometric Payment Guidance}
\label{guidance}
\textbf{Content of guidance.} Participants expressed a desire for better hands-on guidance. Older adults faced operational challenges both in learning to use biometric payment systems and in knowing how to opt out, which limited their autonomy in using biometric payment (Sec. \ref{Lack of Usage Knowledge}). As P13 (F, 67, non-user) said, \textit{``It would be great to have someone come and teach us how to set it up or use it.''} 

In addition to hands-on guidance, participants also indicated a need for better informational support. Some older adults wanted to understand the potential impact of biometric payment systems before enabling them. 
As P14 (F, 73, non-user) noted, \textit{``If I'm told that using this won't affect other aspects or compromise my privacy, then I would definitely accept it.''} 
Participants also hoped to understand how the biometric payment system actually works. As P18 (F, 50, user) explained,
\textit{``If I'm going to enable something, I'll look into how it works such as what company is behind it and what method it uses.''} 

Some older adults also indicated a desire to better understand the usage scope of biometric payment. P03 (F, 55, non-user) explained, \textit{``I need to know the usage scope. If it's only used occasionally, I'd rather use old payment ways.''} 

\textbf{Formats of guidance.}
Participants also shared preferences regarding how guidance could be delivered. First, in-person instruction from professionals was favored over phone-based support. As P06 (F, 61, user) explained, \textit{``It's better to go in person, like to the \modi{ICBC}\footnote{\modi{Industrial and Commercial Bank of China. A Chinese partially state-owned multinational banking and financial services corporation.}}, which is more trustworthy...If professionals can come and teach us how to disable biometric payment or explain how to use it.''} 

Second, peer support emerged as a potentially promising form. P11 (F, 67, non-user) described, \textit{``Our children don't have time, and even when we ask, they're not always willing to explain. So we go back to the senior university. Friends and classmates learn together. In a group, usually one or two people are more tech-savvy, and we learn from them.''} This implies that peer support might be an accessible and acceptable form of guidance for older adults, potentially fostering a more supportive learning environment. 

Finally, while not explicitly framed as a request, one participant implicitly pointed to the need for more accessible learning opportunities in everyday community settings. As P11 (F, 67, non-user) explained, \textit{``But many older people in society or in local communities can't attend senior universities. Very few people actually go to study at those universities. So it's just up to us who have learned something to tell them.''} 

\section{Discussion}
Based on the findings, we discuss three tensions in older adults' experiences with biometric payment (Sec. \ref{tension}), compare our findings with prior work on older adults' usage of general digital payment and biometric technology (Sec. \ref{positioning}), and provide design implications for more inclusive biometric payment systems (Sec. \ref{implications}).

\subsection{Unraveling Biometric Payment Usage through the Lens of Older Adults}
\label{tension}
\textbf{Tension between ease of use and security and privacy concerns.} In Sec. \ref{motivations}, we found that even when older adults were aware of the potential security and privacy risks associated with biometric payment, they still chose to use these systems for their perceived convenience. Prior research has identified a risk-benefit trade-off regarding facial recognition payment in the general population \cite{hu2023should, liu2021resistance}. Our study extends this finding to older adults, suggesting that their decision-making also reflects a similar calculus. 

Interestingly, some older adults did not simply tolerate these potential risks but actively sought to mitigate them, aiming to strike a balance between usability and S\&P protection. For example, as shown in Sec. \ref{protective behaviors}, some participants reported using protective measures to reduce their potential financial loss. These behaviors reflect deliberate efforts to manage the inherent trade-offs in biometric payment, rather than simply accepting potential security and privacy risks for the sake of convenience. Such proactive efforts align with insights from the Protection Motivation Theory (PMT) \cite{rogers1975protection}, which highlights how individuals often consider both the potential severity of risks (threat appraisal) and their ability to effectively manage these risks (coping appraisal) when deciding whether to engage with potentially risky technologies. However, it remains unclear how the general public engages in similar protective behaviors, highlighting a future research direction to explore. 



Moreover, this tension aligns with findings from the general Chinese consumer market. For instance, Liu et al. found that in the early stages of facial recognition payment, many Chinese users were willing to accept privacy risks in exchange for perceived benefits \cite{liu2021resistance}. Cultural variations play a significant role in shaping this risk-benefit assessment. This perspective contrasts with findings from studies conducted in the U.S., where biometric privacy is often taken more seriously. Kugler noted that many American users are so concerned about biometric privacy that they are even willing to forgo tangible benefits to protect their privacy \cite{kugler2019identification}. \modi{One possible explanation is that Chinese users tend to prioritize convenience and functionality over privacy compared to users in Western countries, as shown in prior studies on technology use \cite{lowry2011privacy,liu2021resistance,cao2008user}, and their collectivist culture makes them more cooperative when responding to new technological trends \cite{johnston2009national}.}
Future research can explore potential cultural differences in this risk-benefit tension and how it might affect subsequent protective behaviors. 

\textbf{Tension between password-free benefits and perceived lack of control.}
We found that the password-free nature of biometric payment, typically considered a benefit, can actually lead some older adults to feel a lack of control (Sec. \ref{lack of control}). While prior research and media coverage often frame the removal of password input as a senior-friendly feature that could help mitigate age-related declines (e.g., in vision or memory) \cite{iqbal2020novel,url6}, our findings suggest that this approach may have unintended psychological effects for some older adults. 
\modi{This perception is closely related to the \textit{perceived behavioral control} component of the Theory of Planned Behavior (TPB) that emphasizes attitudes, subjective norms, and \textit{perceived behavioral control} in shaping intention \cite{ajzen1991theory}. In particular, the \textit{perceived behavioral control} component posits that an individual's intention to engage in a behavior is partly determined by their belief in their ability to control the behavior}. In this context, the absence of a password removes a critical and deliberate step that some older adults rely on to confirm their intent, potentially undermining their confidence in managing financial transactions.

Additionally, the physical and cognitive declines that often accompany aging, such as deteriorating eyesight, can further exacerbate this loss of control, as shown in Sec. \ref{lack of control}. 
While removing password requirements reduces the cognitive load and effort involved, it also eliminates this critical opportunity for intentional confirmation, potentially leaving some older adults uneasy about the process. This finding suggests that current biometric payment systems may not fully consider the psychological needs and concerns of older users. Therefore, future biometric payment systems should consider how to balance the benefits and sense of control.


\textbf{Tension between perceptions and technical realities.} 
In Sec. \ref{Lack of Usage Knowledge}, we found that some older adults, due to a lack of knowledge, may hold misconceptions that do not reflect the reality of biometric payment systems. 
\modi{For example, some believed that the usage scope of biometric payment was restricted to a single and specific store. 
Such misconceptions could negatively impact \textit{perceived usefulness} as a critical component of the Technology Acceptance Model (TAM) \cite{davis1989perceived}, potentially discouraging greater adoption and preventing older adults from fully benefiting from its convenience.} Future work could explore how these misconceptions occur and how to educate older adults about the true capabilities and limitations of biometric payment systems.

Moreover, it remains uncertain whether the concerns expressed by older adults in Sec. \ref{concerns} accurately reflect the actual risks involved or if some of these concerns are potentially overestimated. For instance, due to the experience with A4-sized printed photos potentially spoofing facial recognition machines at the workplace, one older adult assumed that the same level of security risk applied to biometric payment systems (Sec. \ref{sp concerns}). This perception might not align with the actual technological architecture and risk profiles of these systems, suggesting a potential gap between user perception and reality. Additionally, we observed in Sec. \ref{protective behaviors} that some older adults avoid video calls to prevent facial data leakage. However, without expert validation, it is difficult to determine whether these strategies are genuinely effective, necessary, or merely based on misconceptions about the risks. Therefore, future work should integrate expert perspectives to assess the legitimacy of older adults' concerns and the effectiveness of their protective behaviors. 

\subsection{Positioning Biometric Payment in Older Adults' Broader Technology Use}
\label{positioning}

\textbf{Digital payment vs. biometric payment: shared foundations and new complexities.} Older adults' motivations for using biometric payment in our findings align closely with those in prior studies on older adults' digital payment. Specifically, corresponding to Sec. \ref{motivations}, convenience \cite{omotayo2020digital}, perceived security \cite{hanafi2020influences}, social influence \cite{jin2022used}, and facilitation conditions \cite{jin2022used} have all been identified as key motivators for older adults across digital payment contexts. 
\modi{These motivators align with the core constructs of the Unified Theory of Acceptance and Use of Technology (UTAUT) respectively, which includes performance expectancy, effort expectancy, social influence, and facilitating conditions as the key elements of accepting and using new technologies} \cite{venkatesh2003user}.

Nevertheless, biometric payment introduces additional perceived vulnerabilities and amplifies existing digital payment concerns.
The perceived uncertainty of the technology and intangibility of provided services have been identified as issues in older adults' interactions with mobile payment~\cite{yang2015understanding}. 
The use of biometric authentication without requiring passwords may further intensify older adults' perceived lack of control over the transaction process, as shown in Sec. \ref{lack of control}. 
Similarly, while concerns around security and privacy have long existed in digital payment \cite{omotayo2020digital,jin2022used}, biometric payment introduces additional and biometric-based concerns, such as biometric data leakage and hacking via biometric spoofing in Sec. \ref{sp concerns}. 
Future research aiming to understand biometric payment or guide system design should recognize these unique and biometric-specific aspects to better support older adults and promote more inclusive digital financial experiences. \modi{It is also worth noting that older adults may have a preference for one biometric payment method over another (e.g., some favored fingerprint payment over facial recognition payment because its deliberate action gives them a sense of control, as shown in Sec. \ref{lack of control}). However, we did not conduct a systematic comparison due to the limited number of participants who have used multiple biometric payment methods. We call for future work to systematically examine user perceptions across different biometric payment methods.}

\textbf{Biometric technology vs. biometric payment: perceived risks carry across contexts.} Biometric technology is used not only in payment scenarios (i.e., biometric payment), but also in a wide range of non-payment contexts such as door access control \cite{kanagamalliga2025biometric} and employee attendance systems \cite{wati2021security}. Our findings suggest that older adults' experiences with biometric authentication in these non-payment settings can influence how they perceive biometric payment systems. Specifically, as described in Sec. \ref{sp concerns}, some participants’ concerns about hacking via biometric spoofing in payment systems stemmed from prior negative experiences with biometric authentication in other domains. They feared that if biometric systems are vulnerable in non-financial settings, the same weaknesses could be exploited to compromise payment security. 
These perceptions indicate that older adults' concerns in one domain may transfer to another, undermining trust in biometric payment and discouraging its adoption. This spill-over effect of perceived risks calls for greater clarity in system design and user education. Future systems could explicitly inform users whether biometric payment systems are isolated from other applications or databases, to help alleviate unnecessary anxieties. 

\subsection{Toward Inclusive Biometric Payment}
\label{implications}

\textbf{Combining biometric and cognitive authentication.} In Sec. \ref{tension}, we discussed the tension between the password-free benefit and older adults' perceived lack of control in biometric payment. To better balance them, older adults envisioned various potential improvements in Sec. \ref{cognitive confirmation}. Although these methods vary in form, they all serve the same fundamental purpose, i.e., providing a cognitive confirmation process in addition to biometric recognition that allows users to consciously verify their intent before completing a payment transaction. 

Moreover, older adults' expected cognitive confirmation mechanisms have several key characteristics. First, they tend to feature low cognitive load, suggesting that the design should strike a balance between security and usability and ensure that the cognitive burden remains manageable. Second, many of these mechanisms incorporate personalization. \modi{While personalized gestures and visual patterns strengthen users' sense of control, this approach introduces new usability challenges, as personalized codes might need regular updates and reliable recall. The implementation needs to understand how often users are willing to change these inputs, how often such patterns must be updated, whether users can manage this effort, and how to effectively integrate biometric recognition with personalized codes to achieve a balance between low cognitive load and a strong sense of control.} A third characteristic is context awareness. It can reduce the risk of unintended transactions by linking the verification process directly to the specific context of each transaction \modi{such as the transaction amount and items. However, it may raise privacy concerns in public settings about the leakage of transaction information.} In addition, these methods also need to account for the diverse physical and cognitive abilities of older adults. For example, gesture-based contextual inputs might be more accessible for some users, while voice-based confirmations might be preferable for those with limited mobility or dexterity.



\modi{Beyond these, their feasibility in real-world payment environments requires further examination. Many public payment terminals, such as those in small retail stores or bank self-service kiosks, operate with heterogeneous hardware capabilities, which may not support fine-grained facial action detection, gesture recognition, or contextual voice input. Environmental factors such as noise or low lighting can further undermine the accuracy of these modalities. These constraints highlight the need to stress-test such mechanisms across representative usage scenarios.}

Overall, these ideas can also serve as design prompts to engage older adults in co-designing confirmation mechanisms that align with their needs. Future research could leverage these insights to collaboratively develop and evaluate different confirmation methods, identifying approaches that could enhance older adults' sense of control and security.





\textbf{Supporting older adults' autonomy in adoption.} 
Older adults have voiced a desire for both operational and informational guidance to enhance their autonomy in using biometric payment systems (Sec. \ref{improvements}). Prior research suggests that older adults tend to be more sensitive to loss aversion~\cite{jin2022used,kahneman2011thinking} and often face greater difficulties when using new technologies compared to younger users \cite{jiang2016generational,choudrie2018understanding,yusif2016older}. As such, providing sufficient operational guidance is crucial for building their usage confidence and autonomy. 

One potential design direction is to incorporate interactive and step-by-step instructional features within biometric payment systems, which provide real-time and hands-on support as users navigate unfamiliar processes. For instance, a potential design could involve a tutorial AI agent that allows older users to request guidance, parse their specific needs, and guide them through each step of some tasks, e.g., guiding older adults through biometric payment simulations, to reduce uncertainty and build user confidence. However, given the sensitive nature of biometric payment, these tutorial interactions should be limited to learning scenarios, ensuring that real transactions are only performed independently. 

For informational guidance, tutorials could pay attention to addressing two key aspects. First, they need to fill knowledge gaps by providing clear and comprehensive information such as the potential impact of the biometric payment system and how the system operates as shown in Sec. \ref{guidance}. Given that older adults might have broader knowledge gaps, integrating an AI-driven chatbot with a well-maintained biometric payment knowledge base could offer continuous and on-demand access to accurate information. Second, correcting common misconceptions is a significant aspect. We summarized key misconceptions in \ref{tension}, which can serve as starting points, such as clarifying the actual usage scope of biometric payments. One practical approach could be collecting common misconceptions and presenting accurate information directly on biometric payment devices to correct misunderstandings. 

\textbf{Integrating protections into existing infrastructure.} In Sec. \ref{protective behaviors}, we observed that older adults take various protective measures to secure their transactions. However, some often require additional effort, such as proactively selective use of biometric payment based on payment amounts and contacting banks to enable real-time transaction notifications. 

Future biometric payment systems could integrate these protective measures directly into their design, reducing the burden on users while also providing safeguards for those unaware of such security options. For instance, during the initial setup, the system could proactively prompt users to set a spending limit for biometric payment instead of requiring them to configure it separately. Currently, most mainstream systems focus on encouraging enrollment, with little emphasis on configuring safeguards. Real-time multimodal payment notifications could also be built into the biometric payment system itself, such as immediate post-transaction voice and text alerts, ensuring that older adults are immediately informed about their transactions. This might reduce user effort while democratizing security measures for less tech-savvy users. These design implications could possibly generalize to broader digital payment ecosystems, particularly in addressing the needs of aging populations.

\section{Conclusion}
Our work makes the first attempt to explore older adults' perceptions and practices regarding biometric payment. To achieve this, we conducted semi-structured interviews with 22 older adults in China. Our findings reveal that convenience, security, social influence, and facilitating conditions are key motivators for using biometric payment. However, older adults expressed various concerns, including lack of control, security and privacy risks (e.g., unauthorized biometric data collection and biometric data leakage), adaptability challenges, and lack of usage knowledge. To balance these benefits and risks, many older adults actively adopt protective measures, such as limiting biometric data disclosure. They also identified possible improvements, including the integration of lightweight, personalized, and context-aware cognitive confirmation mechanisms and more comprehensive educational support. Based on these findings, we discuss three key tensions: the balance between ease of use and security and privacy concerns, the password-free benefit versus lack of control, and the gap between perceived understanding and practical reality. We also position biometric payment within the broader contexts of digital payment and biometric technology and provide recommendations to enhance user control, empower autonomy, and facilitate protective measures.

\begin{acks}
The research is partially funded by the Deutsche Forschungsgemeinschaft (DFG, German Research Foundation) under Germany’s Excellence Strategy - EXC 2092 CASA - 390781972, NSFC grant 62432008, RGC RIF grant R6021-20, RGC TRS grant T43-513/23N-2, RGC CRF grant C6015-23G, NSFC/RGC grant CRS\_HKUST601/24, and RGC GRF grants 16207922, 16207423 and 16203824. We thank Lin Kyi for her valuable feedback on our early drafts. 
\end{acks}

\bibliographystyle{ACM-Reference-Format}
\bibliography{sample-base}
\appendix
\newpage
\section{Interview Guide}\label{interview guide}

\begin{enumerate}
    \item \textbf{Basic Understanding of Biometric Payment}
\begin{enumerate}
    \item Have you heard of biometric payment before?
    \item Where did you first hear about it?
    \item Have you used biometric payment before?
\end{enumerate}

\textbf{if users:}
\begin{enumerate}
    \setcounter{enumii}{3}
    \item What type of biometric payment have you used?
    \item Why did you choose this specific type of biometric payment over others?
    \item How often do you use biometric payment?
    \item On which devices have you used biometric payment?
\end{enumerate}

\item \textbf{General Perceptions of Biometric Payment}

\textbf{if users:}

\begin{enumerate}
    \item What do you think about biometric payment?
    \item Can you describe a memorable experience with biometric payment? Or, what is your usual experience like? 
    \item How did it make you feel?
    \item What were your expectations before using biometric payment?
    \item What considerations did you have before trying it?
    \item Has your perception of biometric payment changed since your first use?
\end{enumerate}

\textbf{if non-users:}

\begin{enumerate}
    \item What do you think about biometric payment?
    \item Why haven’t you used biometric payment? What has prevented you from using it?
    \item Are you willing to try it? Why or why not?
    \item If you were willing to try, what considerations would you have beforehand?
    \item What do you think biometric payment would be like?
\end{enumerate}

\item \textbf{Motivations for Using Biometric Payment}

\textbf{if users:}

\begin{enumerate}
    \item Why did you decide to use biometric payment? What motivated you to try it?
    \item Was it recommended by someone else? Were you influenced by others’ usage?
    \item How did you set up your biometric payment method? Did someone assist you? Was it easy to set up?
    \item In what situations would you use or avoid using biometric payment?
    \item Would you recommend biometric payment to others? Why or why not?
    \item Have your friends used it? What was their experience like?
    \item Compared to QR code payment or cash, do you prefer biometric payment (facial recognition, palm recognition, fingerprint)? Why?
\end{enumerate}

\item \textbf{Concerns about Biometric Payment}

\begin{enumerate}
    \item Did you encounter any difficulties while using it/trying to use it? How did you resolve them?
    \item Do you have any concerns about biometric payment?
    \item Have these concerns you mentioned influenced your decision-making? Have they stopped you from using biometric payment or changed how you use it?
    \item What is the source of your concerns about biometric payment? (e.g., general concerns about mobile payments/technology, or specific to biometric payment? environmental factors/information/interactions? personal experience/external information?)
    \item Have you ever experienced or heard of cases where money was stolen, information was compromised, transactions were delayed for a long time, or other negative issues occurred while using biometric payment?
    \item Do you have any security and privacy concerns about biometric payment?
    \item What does security/privacy mean to you in the context of biometric payment?
    \item Regarding security/privacy concerns, which parties do you distrust? (e.g., merchants, banks, regulators, payment services like Alipay/WeChat Pay, and tech companies?)
    \item Who do you think is responsible for your privacy and security in biometric payment?
    \item How do you think your collected biometric information will be used?
    \item How well do you understand biometric payment technology itself? (e.g., how does it recognize or process payment?)
\end{enumerate}

\item \textbf{Protective Behaviors}

\textbf{if users:}

\begin{enumerate}
    \item Do you take specific measures to protect your information or money? 
    \item What steps have you taken to ensure your security and privacy?
    \item Are these measures specific to online payments or biometric payment in particular?
    \item When you make payments, do you usually do it alone or with others? If you are with others, do you have any special ways to protect yourself?
    \item Have these measures changed over time?
\end{enumerate}

\item \textbf{Desired Improvements}

\begin{enumerate}
    \item In an ideal scenario, what kind of biometric payment would you be willing to use?
    \item What functionalities should it have?
    \item What features or designs could address your concerns?
    \item How should the system protect your privacy?
    \item How should the system ensure your security?
\end{enumerate}

\item \textbf{Demographic Questions}

\begin{enumerate}
    \item Gender
    \item Age
    \item Location
    \item Education level
\end{enumerate}
\end{enumerate}
\end{document}